\documentclass[twocolumn,english,showpacs,showkeys,reprint,linenumbers]{revtex4}
\usepackage[T1]{fontenc}
\usepackage[latin9]{inputenc}
\usepackage{textcomp}
\usepackage{amsmath}
\usepackage{graphicx}

\makeatletter
\@ifundefined{textcolor}{}
{%
 \definecolor{BLACK}{gray}{0}
 \definecolor{WHITE}{gray}{1}
 \definecolor{RED}{rgb}{1,0,0}
 \definecolor{GREEN}{rgb}{0,1,0}
 \definecolor{BLUE}{rgb}{0,0,1}
 \definecolor{CYAN}{cmyk}{1,0,0,0}
 \definecolor{MAGENTA}{cmyk}{0,1,0,0}
 \definecolor{YELLOW}{cmyk}{0,0,1,0}
}

\usepackage{babel}

\makeatother

\usepackage{babel}
\begin{document}

\title{Combining ferromagnetic resonator and digital image correlation to
study the strain induced resonance tunability in magnetoelectric heterostructures}

\author{Fatih Zighem$^{1}$}

\email{zighem@univ-paris13.fr}

\selectlanguage{english}%

\author{Mohamed Belmeguenai$^{1}$, Damien Faurie$^{1}$}

\email{faurie@univ-paris13.fr}

\selectlanguage{english}%

\author{Halim Haddadi$^{2}$ }

\author{Johan Moulin$^{3}$}

\affiliation{$^{1}$Laboratoire des Sciences des Procédés et des Matériaux, CNRS-Université
Paris XIII, Sorbonne Paris Cité, Villetaneuse, France}

\affiliation{$^{2}$Laboratoire MSMP\textemdash{}Carnot Arts, ENSAM ParisTech,
rue Saint-Dominique, 51006 Châlons-en-Champagne, France}

\affiliation{$^{3}$Institut d'Electronique Fondamentale, UMR 8622, Université
Paris Sud-CNRS, Orsay, France}

\date{July 28 2014 }
\begin{abstract}
This paper reports the development of a methodology combining microstrip
ferromagnetic resonance (MS-FMR) and digital image correlation (DIC)
in order to silmuteanously measure the voltage-induced strains and
the magnetic resonance in artificial magnetoelectric heterostructures
(``magnetic films/piezoelectric substrate'' or ``magnetic films/flexible
substrate/piezoelectric actuator''). The overall principle of the
technique and the related analytical modelling are described. It is
powerful to estimate the magnetostriction coefficient of ferromagnetic
thin films and can be used to determine the effective magnetoelectric
coefficient of the whole heterostructures in addition to the piezoelectric
coefficient related to the in-plane voltage-induced strains. This
methodology can be applied to system for which the strains are well
transmitted at the different interfaces.
\end{abstract}


\maketitle

\section{Introduction}

During the last decade, many groups have concentrated their effort
on the study of ferromagnetic resonance (FMR) tunability through electric
field strain induced MagnetoElectric (ME) coupling \cite{Fetisov2006,Srinivasan2010}.
Many ME systems based on ferromagnetic/ferroelectric heterostructures
have been developed, encompassing voltage-tunable microwave signal
processing devices \cite{Subramanyam2013,Liu2013,Zavislyak2013},
magnetoelectric random access memory devices\cite{Nan2012,Yang2014,Fusil2014,Jin2014}
and strain-control GMR devices \cite{Rizwan2013,Liu2011,Lei2013}.
For all these kinds of systems, the properties are controlled \textit{via}
the elastic voltage-induced strains transmitted from the ferroelectric
medium to the ferromagnetic one. Hence, the knowledge of the elastic
strains must be known in order to predict the changes of magnetic
state induced by applying voltages. Concerning magnetic thin films
deposited on substrates, there are two ways to control the magnetization
by applying voltage-induced strains : i) deposition of the film on
a piezoelectric substrate \cite{Filippov2008,Thiele2007,Park2010}
and ii) deposition of the film on a ``non-piezoelectric'' substrate
and subsequent cementation on a piezoelectric actuator \cite{Pettiford2008,Brandlmaier2008_PRB,Brandlmaier2008_PRB_Bis}.
In the first one, the in-plane strain transmission is generally good,
the main disadvantage being to transfer existing process to piezoelectric
substrates (PZT, PZN-PT, PMN-PT, ...) ; in the second one, the main
problem is the possibly bad in-plane strain transmission when the
substrate is rigid (Si, GaAs, MgO, ...), the total transmission being
attained only with compliant substrates (polymers) \cite{Zighem_JAP2013}.
In the two cases of full transmission (``film/piezoelectric substrate'',
``film/polymer substrate/piezoelectric actuator''), the knowledge
of the in-plane strains in the piezoelectric substrate or actuator
is sufficient to know the in-plane strains in the whole system. In
order to measure simultaneously the in-plane strains and ferromagnetic
resonance in such systems, we have developed a methodology combining
\textit{in situ} microstrip ferromagnetic resonance (MS-FMR) with
the Digital Image Correlation (DIC) technique. In this paper, we will
show the general principle of the technique including the analytical
formalism, before describing the experimental details. Then we will
show an example of study on a ``Finemet\textregistered{} film/Kapton\textregistered{}
substrate/piezoelectric actuator'' system highlighting the potentialities
of the method. Especially, we will show that this technique is very
efficient to determine the magnetostriction coefficient at saturation
of the film, in complement to recently developed techniques for thin
films, such as magnetoelastic measurement setup with a MOKE magnetometer
\cite{Will2012}, substrate deflection method under magnetic field
\cite{Bruckner2001}, laser Doppler vibrometry \cite{Varghesea2014},
applied whole wafer curvature \cite{Hill2013}, strain modulated ferromagnetic
resonance \cite{Nesteruk2014}, or vibrating sample magnetometer coupled
with bending\cite{Buford2014}.

\section{Principle}

\subsection{Methodology}

The voltage strain-induced magnetoelectric coupling in artificial
multiferroic systems, consisting on a magnetic film deposited onto
a piezoelectric substrate, is based on a two step process: i) the
application of an electric field inside the piezoelectric medium induces
mechanical deformations to the magnetic film \textit{via} piezoelectric
effect and ii) magnetostriction of the magnetic film induces a magnetoelastic
anisotropy which can be strain(or voltage)-tunned in amplitude and
direction. In these conditions, a quantitative evaluation of this
indirect magnetoelectric coupling requires both the determination
of the strains induced by the piezoelectric medium to the ferromagnetic
film and the induced magnetoelastic anisotropy. These parameters can
be evaluated with our experimental setup. Indeed, with our setup,
when applying a voltage to the piezoelectric medium, it is possible
to simultaneously measure the induced mechanical deformation and the
magnetoelastic anisotropy thanks to a combination of digital image
correlation and broadband ferromagnetic resonance techniques. Indeed,
the uniform precession mode resonance field is directly linked to
the magnetoelastic anisotropy when the film is submitted to external
stresses. In these conditions, it is not only possible to measure
an effective magnetoeletric coupling which do not necessary require
the strains evaluation inside the magnetic film but also the magnetostriction
coefficients. The following paragraph presents a basic analytical
modelling of the resonance field variation as function of external
applied stresses.

\begin{figure}
\includegraphics[bb=40bp 280bp 410bp 590bp,clip,width=7cm]{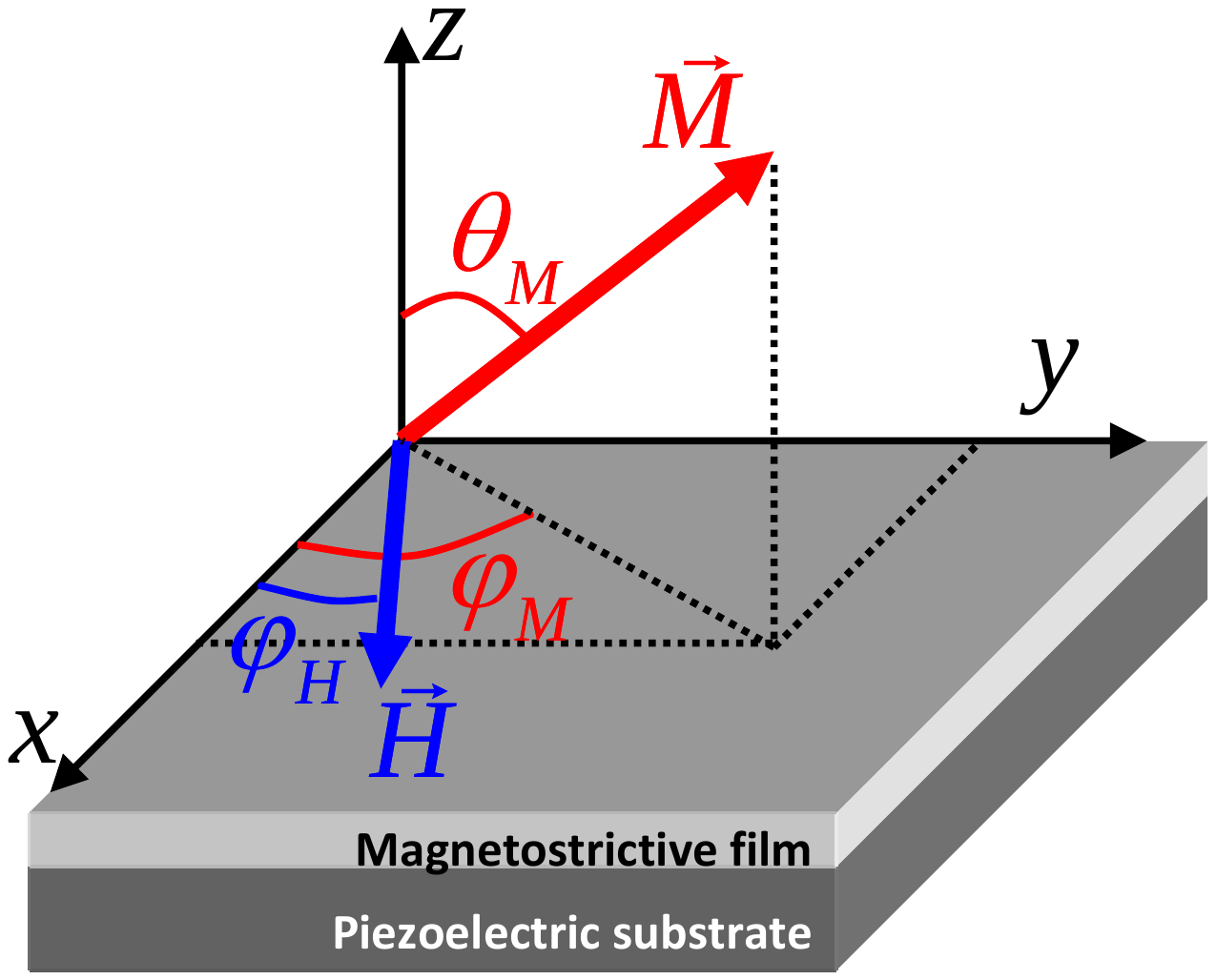}

\caption{Schematic illustration showing angles, fields and coordinate systems
used in the text. }

\label{Fig_Sketch_FMR}
\end{figure}

\subsection{Theoretical background }

In this paragraph, the resonance field of the uniform precession mode
of a magnetostrictive film submitted to in-plane external stresses
is derived in the macrospin approximation (\textit{i. e.} a uniform
magnetization is considered). In our system, these in-plane stresses
are applied by the piezoelectric medium to the ferromagnetic film
(see figure \ref{Fig_Sketch_FMR}). For simplicity, the magnetostrictive
and elastic properties of this thin film are considered as isotropic
(it will be the case in the studied system). With these assumptions
(isotropic behavior and macrospin approximation), the magnetostriction
coefficient, the Young's modulus and the Poisson's ratio of the thin
film are scalars parameters. The magnetic energy of the thin film,
using the coordinates system presented in Figure \ref{Fig_Sketch_FMR}
can be written as:

\begin{equation}
F=F_{zee}+F_{dip}+F_{me}
\end{equation}

Where the two first terms stand for the Zeeman and the dipolar contributions,
respectively. The last term corresponds to the magnetoelastic anisotropy
term and can be written as:
\begin{multline}
F_{me}=-\frac{3}{2}\lambda\Big(\big(\cos^{2}\varphi_{M}\sin^{2}\theta_{M}-\frac{1}{3}\big)\sigma_{xx}\\
+\big(\sin^{2}\varphi_{M}\sin^{2}\theta_{M}-\frac{1}{3}\big)\sigma_{yy}\Big)
\end{multline}

$\sigma_{xx}$ and $\sigma_{yy}$ being the in-plane principal stress
tensor components while $\theta_{M}$ and $\varphi_{M}$ stand for
the polar and the azimuthal angles of the magnetization $\vec{M}$.
Finally, $\lambda$ is the isotropic magnetostriction coefficient
at saturation of the thin film. The relation between the principal
stress components ($\sigma_{xx}$, $\sigma_{yy}$ ) and strains ($\varepsilon_{xx}$,
$\varepsilon_{yy}$) tensors is given by the isotropic Hook's law
where $E$ is the Young's modulus and $\nu$ is the Poisson's ratio:

\begin{flalign}
\sigma_{xx} & =\left(\frac{E}{1+\nu}\right)\left(\frac{1}{1-\nu}\varepsilon_{xx}+\frac{\nu}{1-\nu}\varepsilon_{yy}\right)\label{eq:Hook_law_1}\\
\sigma_{yy} & =\left(\frac{E}{1+\nu}\right)\left(\frac{1}{1-\nu}\varepsilon_{yy}+\frac{\nu}{1-\nu}\varepsilon_{xx}\right)\label{eq:Hook_law_2}
\end{flalign}
 The resonance field of the uniform precession mode evaluated at the
equilibrium is obtained from the following expression:
\begin{equation}
\left(\frac{2\pi f}{\gamma}\right)^{2}=\left(\frac{1}{M_{S}\sin\theta_{M}}\right)^{2}\left(\frac{\partial^{2}F}{\partial\theta_{M}^{2}}\frac{\partial^{2}F}{\partial\varphi_{M}^{2}}-\left(\frac{\partial^{2}F}{\partial\theta_{M}\varphi_{M}}\right)^{2}\right)\label{eq:Smit_Beljers}
\end{equation}

\begin{figure*}
\includegraphics[bb=40bp 110bp 710bp 575bp,clip,width=15cm]{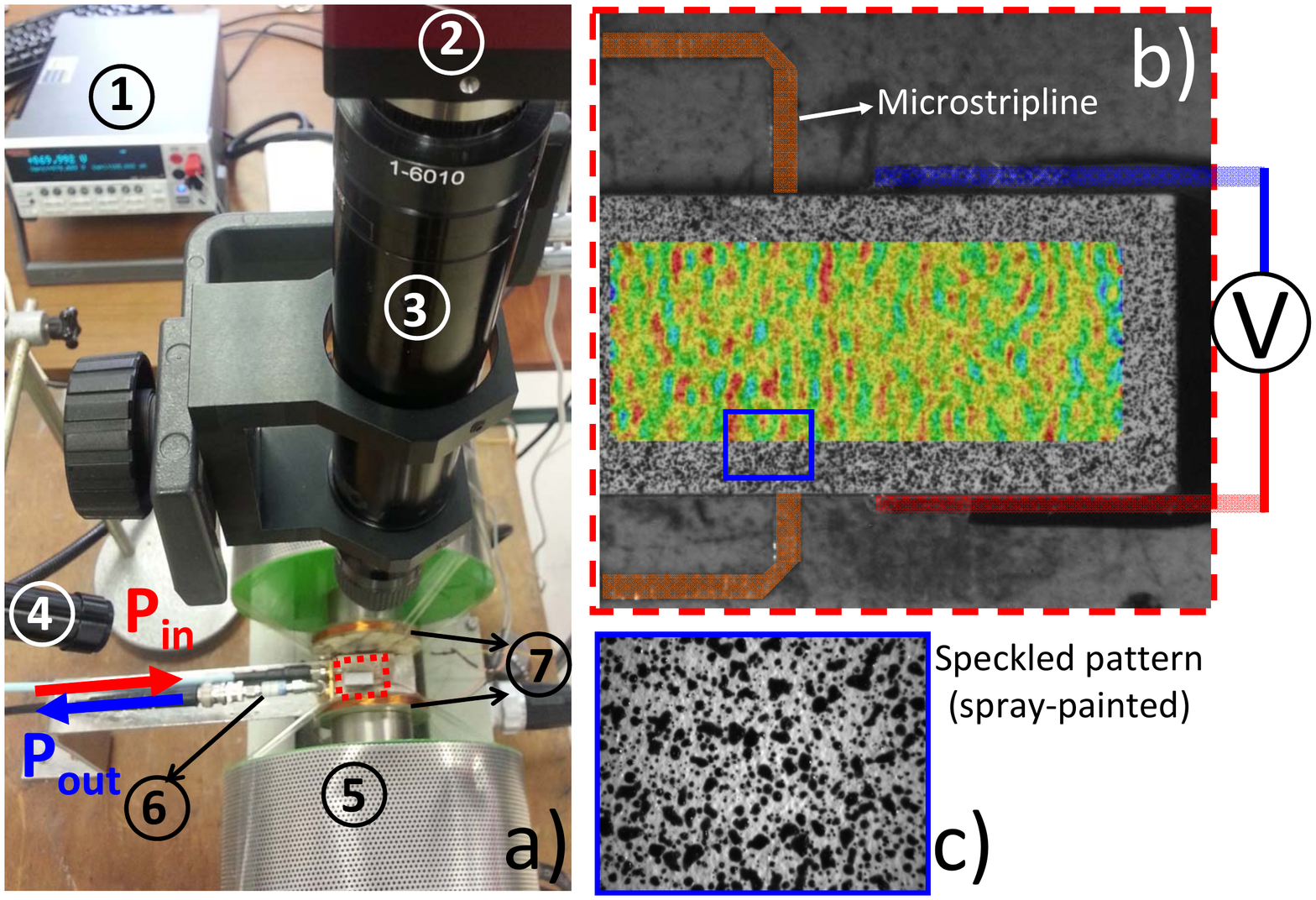}

\caption{a) Setup image of the combined MS-FMR/DIC experiment. The circled
numbers correspond to:  \textcircled {1}: Keithley Model 2400;  \textcircled {2}:
CCD camera (AVT-Pike-f421b); \textcircled {3}: objective lens for
the CCD camera;\textcircled {4}: white light source;\textcircled {5}:
electromagnet;\textcircled {6}: Schottky detector and \textcircled {7}:
modulation coils. $P_{in}$ and $P_{out}$ are the injected and transmitted
radio frequency current. b) Zoom in showing the sample mounted onto
the microstripline. A typical calculated strain field map is present
at the top of the sample. c) Zoom in of the speckled pattern (spray-painted)
at the top of the sample necessary for the strain fields calculations.}

\label{Fig_Setup}
\end{figure*}

Where $f$ is the microwave driving frequency and $\gamma$ is the
gyromagnetic factor ($\gamma=g\times8.794\times10^{6}$ s$^{-1}$.Oe$^{-1}$).
The different energy derivatives are evaluated for the equilibrium
direction of the magnetization. For an in-plane applied magnetic field,
the equilibrium polar angle is $\varphi_{M}=\frac{\pi}{2}$ because
of the large effective demagnetizing field associated with the planar
film geometry and an explicit expression is obtained:

\begin{multline}
\left(\frac{2\pi f}{\gamma}\right)^{2}=\Big[4\pi M_{s}+H_{res}\cos(\varphi_{M}-\varphi_{H})+\\
\frac{3\lambda}{M_{s}}\Big(\sigma_{xx}\cos^{2}\varphi_{M}+\sigma_{yy}\sin^{2}\varphi_{M}\Big)\Big]\times\\
\Big[H_{res}\cos(\varphi_{M}-\varphi_{H})+\frac{3\lambda}{M_{s}}\Big(\sigma_{xx}\cos^{2}\varphi_{M}+\sigma_{yy}\sin^{2}\varphi_{M}\Big]\label{eq:Freq}
\end{multline}

Here $\varphi_{H}$ is the angle between the in-plane applied magnetic
field and $x$ direction (see figure \ref{Fig_Sketch_FMR}). The analysis
can be simplified if the resonance field is larger than the magnetoelastic
anisotropy field ($\vec{H}_{me}=-\vec{\nabla}_{\vec{M}}F_{me}$).
In this condition, the magnetization direction is almost aligned along
the applied magnetic field ($\varphi_{M}\sim\varphi_{H}$). The resonance
field is thus given by:

\begin{multline}
H_{res}=\\
\sqrt{\big(2\pi M_{s}+\frac{3\lambda}{2M_{s}}(\sigma_{xx}\sin^{2}\varphi_{H}+\sigma_{yy}\cos^{2}\varphi_{H})\big)^{2}+\Big(\frac{2\pi f}{\gamma}\Big)^{2}}\\
-2\pi M_{s}-\frac{3\lambda}{4M_{s}}\Big(\sigma_{xx}(1+3\cos2\varphi_{H})+\sigma_{yy}(1-3\cos2\varphi_{H})\Big)\label{eq:Resonance_field}
\end{multline}

The two first terms represents an almost constant shift in the resonance
field baseline because $2\pi M_{s}$ and $\frac{2\pi f}{\gamma}$
are found to be larger than the magnetoelastic and the magnetoelastic
anisotropy field. The last term correspond to the angular variation
of the resonance field due to the voltage induced magnetoelastic anisotropy
field.

\section{Experimental Setup}

The experimental setup presented in figure \ref{Fig_Setup} has been
developed in order to \textit{in situ} study the indirect magnetoelectric
effect occurring in artificial magnetoelectric multiferroics heterostructures
such as the ones presented in introduction. Our setup combines microstrip
ferromagnetic resonance (MS-FMR) and digital image correlation (DIC)
techniques. The MS-FMR characterization is performed with the help
of a field modulated FMR setup using HP 83752B generator, operating
in the 0.01-20 GHz frequency range, to generate a radio frequency
field ($\vec{h}_{rf}$) of variable power (-100 dBm to +20 dBm). The
sample (figure \ref{Fig_Setup}b)) is mounted on a 0.5 mm microstrip
line (the film side is in direct contact with the microstripline)
connected to the HP generator \textit{via} a semi-flexible SMA cable
and to a lock-in amplifier (Stanford research system SR830) to derive
the field modulated measurements \textit{via} a Schottky detector.
The microstrip line (MS), composed of 0.5 mm Cu-strip grown on Cu-back
side metalized Al$_{2}$O$_{3}$ substrate, is designed to have 50
Ohm characteristic impedance and broadband. The MS and the sample
are inserted inside the gap of an electromagnet connected to a DC
power supplier to generate in-plane magnetic fields up to 5 kOe with
a resolution of 0.1 Oe. This external magnetic field is modulated
at a frequency of 170 Hz with an amplitude varying from 1 to 8 Oe
allowing lock-in detection to be used in order to increase the signal-to-noise
ratio. This broadband MS-FMR offers a high sensitivity allowing to
detect a net magnetic moment down to $10^{-5}$ emu.

\begin{figure}
\includegraphics[bb=20bp 340bp 430bp 585bp,clip,width=8.5cm]{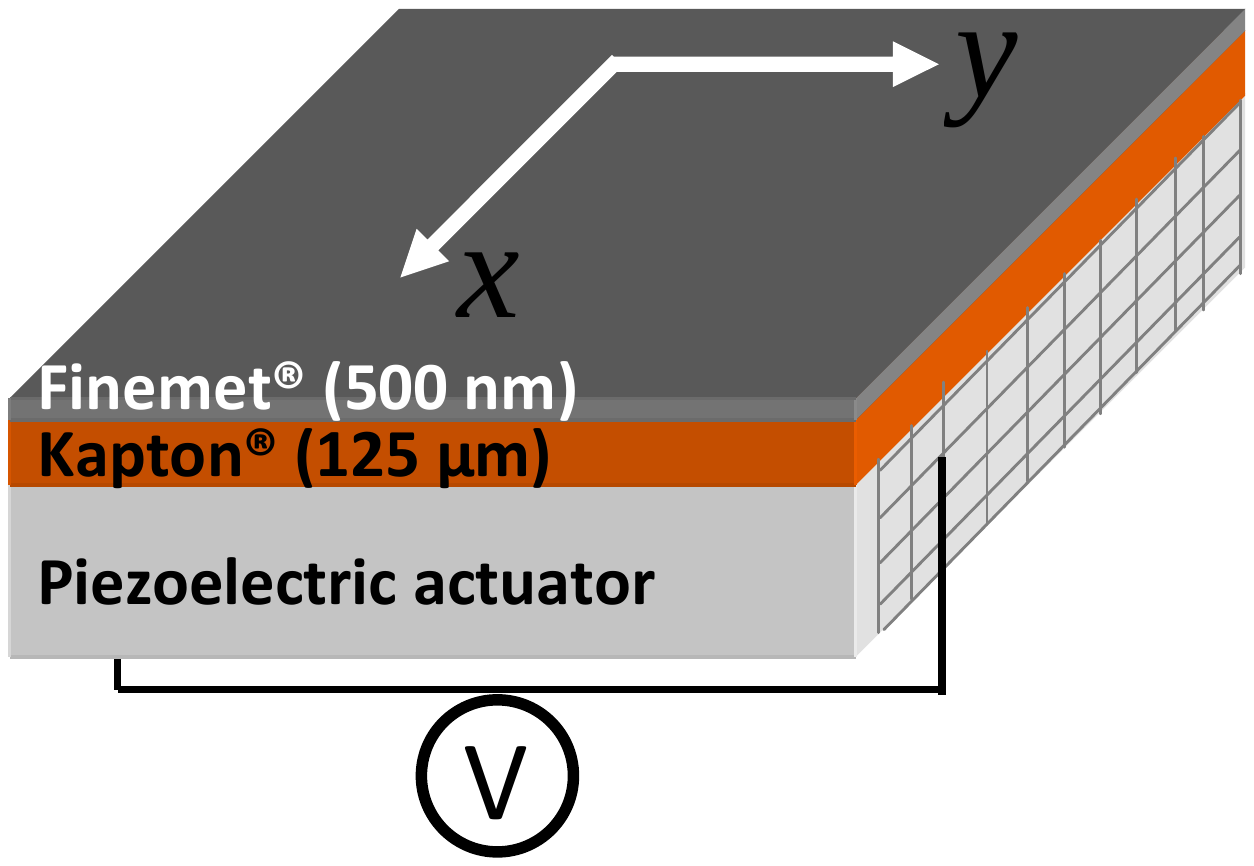}

\caption{Sketch of the studied heterostructure showing the 500 nm thick Finemet\textregistered{}
film deposited onto a 125 $\text{\textmu}$m thick Kapton\textregistered{}
substrate and glued onto the piezoelectric actuator.}

\label{Fig_Sketch_System}
\end{figure}

The piezoelectric media is connected to a Keithley power supplier
(Model 2400) allowing to apply DC voltages in the range {[}-200 V;+200
V{]} with 0.001 V resolution. For each applied voltage, a MS-FMR spectrum
and an image of the top surface of the sample are recorded (note that
the spectrum recording takes from 1 minutes to several hours depending
on the magnetic field step and on the total width of the spectrum).
The images recording at each step will serve to determine the in-plane
field strains. Furthermore, being given the initial ``homogeneous''
surface of the sample, a speckle pattern has been spray painted in
order to generate a contrast which will serve to calculate the strain
fields (an image of a typical speckled pattern (spray-painted) is
presented in Figure \ref{Fig_Setup}c)). In our setup, the images
are recorded thanks to a $2048\times2048$ pixels CCD AVT-Pike f421b
camera vertically positioned in the top of the surface sample. The
objective lens has been chosen in order to be sufficiently far from
the electromagnet (20 cm) with a field view around $1\times1$ cm$^{2}$
as shown on Figure\ref{Fig_Setup}b). The strain fields calculations
(from the different images) have been perfumed by digital image correlation.
Several commercial and open source softwares based on digital image
correlation technique are available (Ncorr \cite{Ncorr}, Moire \cite{Opticist},
Aramis \cite{Aramis}, Matchid \cite{Matchid}, Correli \cite{Correli},
...) for the determination of strains fields. In the following illustrative
results, the DIC calculations have been performed by using Aramis.
Finally, it is worth to mention that this setup is piloted \textit{via}
a Labview program providing flexibility of a real time control of
the magnetic field and the DC voltage sweeps, step and rate, real
time data acquisition and visualization.

\section{Illustrative results}
\begin{figure*}
\includegraphics[bb=20bp 10bp 800bp 570bp,clip,width=13cm]{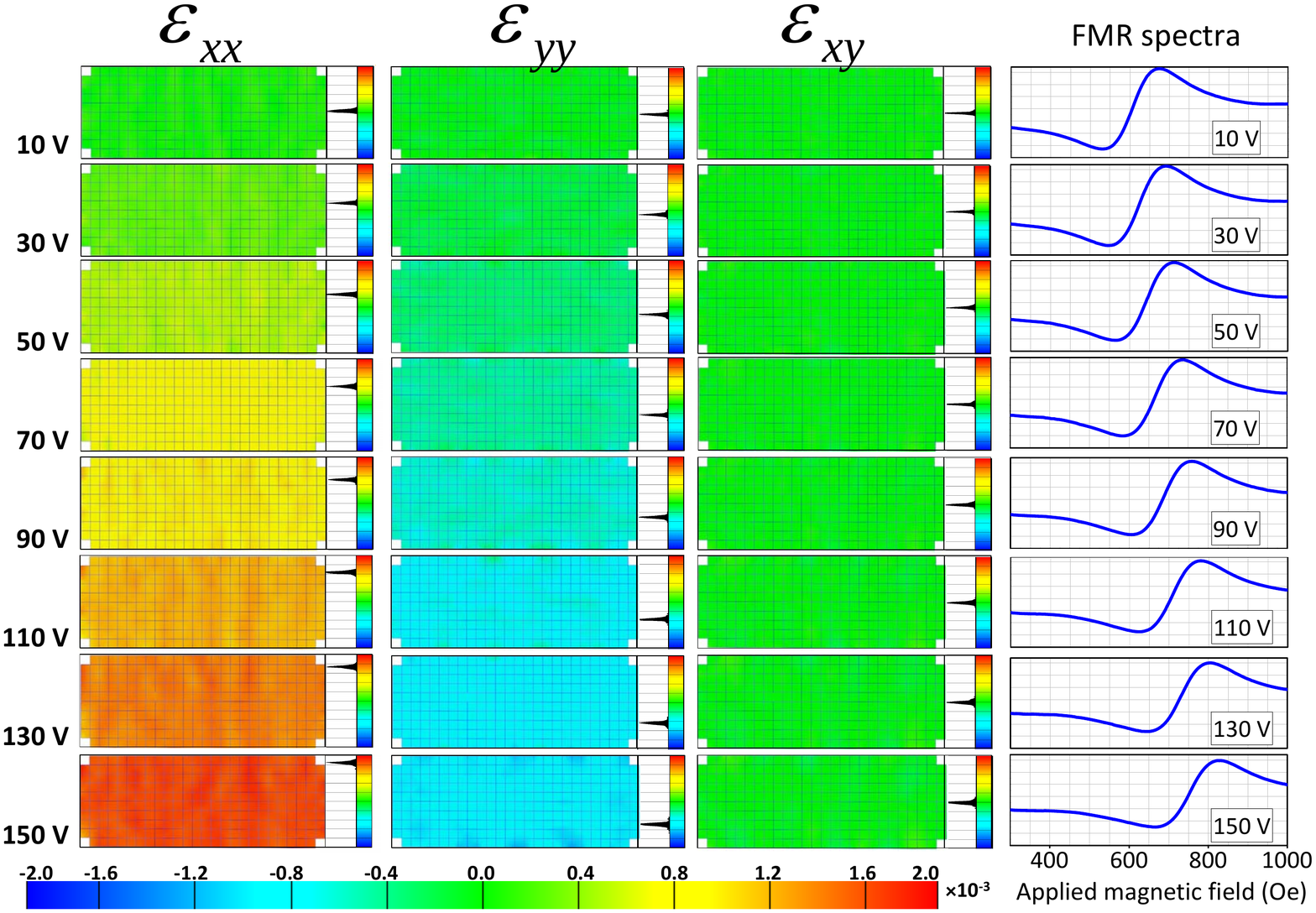}

\caption{Calculated in-plane strains fields ($\varepsilon_{xx}$ , $\varepsilon_{yy}$
and $\varepsilon_{xy}$) and corresponding FMR spectra for specific
applied voltages (from 0 V to $+150$ V). Histograms of the in-plane
strains fields are represented for each applied voltage. Note the
almost zero value of the shear strain ($\varepsilon_{xy}$).}

\label{Fig_Carto}
\end{figure*}
\begin{figure}
\includegraphics[bb=30bp 100bp 660bp 570bp,clip,width=8.5cm]{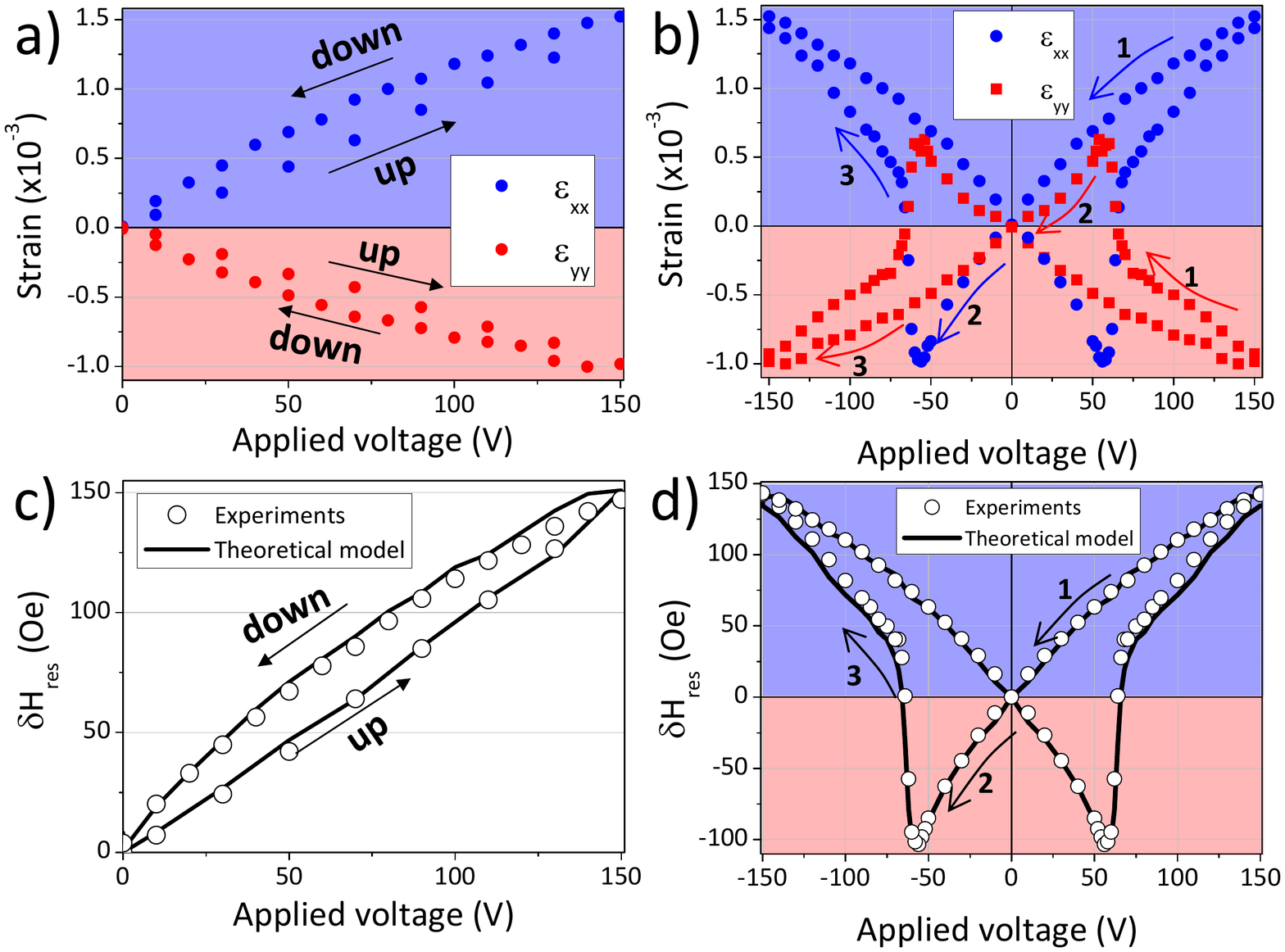}

\caption{Mean in-plane strains ($\varepsilon_{xx}$ and $\varepsilon_{yy}$)
and $\delta H_{res}$ variations as function of the applied voltage
(a,c: from 0 to $+150$ V and back to 0 V; b,d: from $+150$ V to
$-150$ V and back to $+150$ V). The continuous lines in c) and d)
are calculated by using equation \ref{eq:Resonance_field} with the
following parameters: $\gamma=1.885\times10^{7}$ s$^{-1}$.Oe$^{-1}$,
$M_{S}=965$ emu.cm$^{-3}$, $E=145$ GPa and $\nu=0.27$ and $\lambda=17.5\times10^{-6}$.}

\label{Fig_Papillon}
\end{figure}

The methodology is illustrated with a ``magnetic film/polymer substrate/piezoelectric
actuator'' system. Figure \ref{Fig_Sketch_System} presents a cross-section
sketch of the fabricated heterostructure. An amorphous 530 nm thick
Finemet\textregistered{} film has been deposited onto a 125 $\mu$m
thick polyimide flexible substrate (Kapton\textregistered{}) by radio
frequency sputtering. The deposition conditions were a residual pressure
of around 10$^{-7}$ mbar, a working Ar pressure of 40 mbar and a
RF power of 250 W. Prior to the Finemet\textregistered{} deposition,
a 10 nm thick Ti buffer layer , was deposited on the substrate to
ensure a proper adhesion of the Finemet\textregistered{} film. Finally,
another 10 nm thick Ti cap layer was deposited on the top of the Finemet\textregistered{}
film in order to protect it from oxidation. The composition of the
film has been measured by EDS (Energy Dispersive Spectroscopy) and
is close to that of the target (Fe$_{73.5}$Cu$_{1}$Nb$_{3}$Si$_{15.5}$B$_{7}$)
while the thickness of the film (530 nm) has been measured by Scanning
Electron Microscopy. After deposition, the film/substrate system has
been glued onto a piezoelectric actuator.

\begin{figure}
\includegraphics[bb=30bp 340bp 660bp 590bp,clip,width=8.5cm]{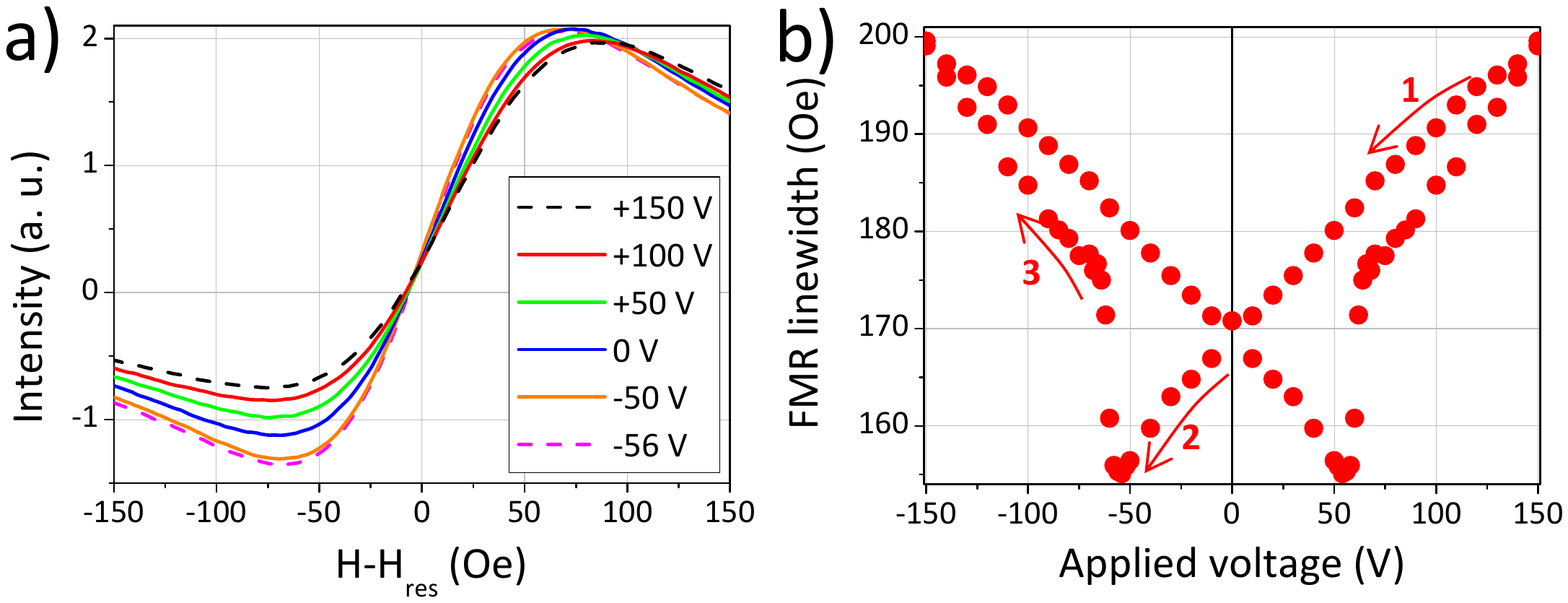}

\caption{a) Typical FMR spectra obtained at 8 GHz showing the linewidth variation
as function of the applied voltage. For clarity, they are centered
around 0 Oe (shifted by their respective resonance field). b) Corresponding
FMR linewidth ($\Delta H_{pp}$) variation as function of the applied
voltage (from $+150$ V to $-150$ V and back to $+150$ V). }

\label{Fig_Linewidth}
\end{figure}

The figure \ref{Fig_Carto} shows the map of the in-plane strains
($\varepsilon_{xx}$, $\varepsilon_{yy}$, $\varepsilon_{xy}$) for
a few voltages applied to the piezoelectric actuator (from 0 V to
$+150$ V). For the three in-plane strain components, the whole distribution
is estimated to be about $10^{-4}$ so that the strain heterogeneities
in the piezoelectric actuator are relatively weak. Obviously, to interpret
the FMR results, we use the mean values of the elastic strains. When
we apply a voltage, the mean values of $\varepsilon_{xx}$ and $\varepsilon_{yy}$
vary with $\frac{\varepsilon_{yy}}{\varepsilon_{xx}}\sim-0.65$, while
the in-plane shear strain $\varepsilon_{xy}$ remains unchanged. This
behavior is illustrated in figures \ref{Fig_Papillon}-a and \ref{Fig_Papillon}-b
where the mean strains $\varepsilon_{xx}$ and $\varepsilon_{yy}$
are plotted as function of the voltage.

In \ref{Fig_Papillon}-a, the curve corresponds to a simple electric
loading-unloading (0V - 150V - 0V) ; the non-linearity of the curve
is due to the specific piezoelectric behavior of the actuator, which
is reversible. In figure \ref{Fig_Papillon}-b, we show an unloading
from 150V to -150V and a subsequent loading from -150V to 150V. We
observe the so-called ``butterfly'' behavior that is due to polarization
switching at about -60V during unloading and 60V during loading. In
addition, we have checked that the analysis of $\varepsilon_{xx}$
and $\varepsilon_{yy}$ on the thin film gives same mean values\cite{Zighem_JAP2013}.
Thus, the transmission of the in-plane strains between the actuator
and the surface of the film is close to $100\%$.

The voltage induced anisotropy has been studied in a specific configuration
(magnetic field $\overrightarrow{H}$ applied along $y$ axis). The
influence of the applied voltage on the magnetic properties of the
thin film has been probed by MS-FMR technique. Figures \ref{Fig_Papillon}-c
and \ref{Fig_Papillon}-d show MS-FMR results obtained from experimental
spectra recorded at 8 GHz with an applied magnetic field along the
$y$ axis (\textit{i. e.}$\varphi_{H}=\frac{\pi}{2}$) and for different
applied voltages; typical experimental spectra are shown on figure
\ref{Fig_Linewidth}-a. In these conditions, the deduced resonance
fields are in a magnetic saturating regime so that we can deduce here
the magnetostriction coefficient at saturation. It clearly appears
that the resonance field increases with the applied voltage, which
indicates that the $y$ axis is harder and the $x$ axis easier for
the magnetization direction when a positive voltage is applied. This
is consistent with a positive magnetostriction coefficient, as expected
for amorphous Finemet\textregistered{} alloys \cite{Moulin2011}.
The shift of the resonance fields as function of the applied voltage
are presented in figures \ref{Fig_Papillon}-c and \ref{Fig_Papillon}-d
and corresponds to the tests shown respectively in figures \ref{Fig_Papillon}-a
and \ref{Fig_Papillon}-b for the strain analysis. Obviously, the
observed non-linear and hysteretic variations of the resonance field
is due to the variations of $\varepsilon_{xx}$ and $\varepsilon_{yy}$
as function of the applied voltage (see figure \ref{Fig_Sketch_System}).
In order to quantitatively bind the resonance field variations to
the in-plane strains induced by the applied voltage, we can calculate
the uniform precession mode frequency as function of the applied voltage
(strain) by adding a magnetoelastic energy term $F_{me}$ to the total
magnetic energy density $F$ of the film, as described previously.
The solid lines in Figures \ref{Fig_Papillon}-c and \ref{Fig_Papillon}-d
are fits to the experimental data calculated by using the parameters
previously determined by MS-FMR ($\gamma=1.885\times10^{7}$ s$^{-1}$.Oe$^{-1}$,
$M_{S}=965$ emu.cm$^{-3}$) and Brillouin light scattering ($E=145$
GPa and $\nu=0.27$) \cite{Fillon2014}. Actually, the magnetostriction
coefficient of the thin film is the sole parameter to be determined
using this experiment and has been estimated to be $\lambda=17.5\times10^{-6}$,
slightly lower that the bulk counterpart ($\lambda=23\times10^{-6}$)
\cite{Moulin2011}.

In addition, in figure \ref{Fig_Linewidth}-b, the FMR peak to peak
linewidth ($\Delta H_{pp}$), defined as the field difference between
the extrema of the sweep-field measured FMR spectra, is plotted as
function of the applied voltage using 8 GHz driving frequency. This
figure shows that $\Delta H_{pp}$ is significantly enhanced by the
applied voltage and interestingly presents similar behavior as the
resonance field shift and the strain (\ref{Fig_Papillon}-b and \ref{Fig_Papillon}-d)
suggesting its correlation with the voltage induced strain. The modelling
of the strain behavior of the FMR linewidth is out of this paper scope
and will be addressed in forthcoming papers. However, this strain
tuning of the FMR linewidth remains a simple and promising manner
to artificially choose the desired damping depending on the aimed
application.

\section{Conclusion and perspectives}

We have shown a methodology that combines microstrip ferromagnetic
resonance (MS-FMR) and digital image correlation (DIC) in order to
study the voltage-induced strains effect on magnetic anisotropy in
thin films. The elastic strains are measured on the actuator (or substrate)
side while the magnetic resonance field is measured in the thin films.
This technique allows determining the magnetostriction coefficient
of the film and can also be used to determine the effective magnetoelectric
coefficient of the whole system and the piezoelectric coefficient
related to the in-plane voltage-induced strains. This methodology
can be applied to system for which the strains are well transmitted
at the different interfaces (``film/piezoelectric substrate'' system
or ``film/polymer substrate/piezoelectric actuator'' system). Moreover,
the strain tuning of the FMR linewidth is promising for spintronics
applications. Indeed, the magnetic damping controls how fast the magnetization
reverses and therefore is interesting technological parameter.
\begin{acknowledgments}
The authors gratefully acknowledge the CNRS for his financial support
through the ``PEPS INSIS'' program (FERROFLEX project) and the Université
Paris 13 through a ``Bonus Qualité Recherche''. \end{acknowledgments}

\end{document}